# PAPERS & REVIEWS

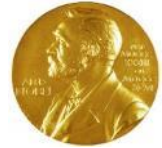

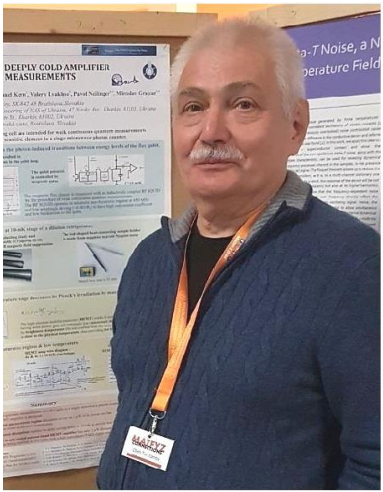

**Oleh G. TURUTANOV —**
Candidate of Sciences in Physics and Math (PhD),
Senior researcher in the Dept. of Superconducting and Mesoscopic Structures of B.I. Verkin Institute for Low Temperature Physics and Engineering of the NAS of Ukraine,
Senior researcher in the Dept. of Experimental Physics at the Faculty of Mathematics, Physics and Informatics at Comenius University Bratislava (Slovakia)

ORCID:
https://orcid.org/0000-0002-8673-136X

## THE NOBEL PRIZE IN PHYSICS AND THE CONTRIBUTION OF UKRAINIAN SCIENTISTS TO THE UNDERSTANDING OF QUANTUM PHENOMENA, IN PARTICULAR THE BEHAVIOR OF MACROSCOPIC QUANTUM SYSTEMS
### The 2025 Nobel Prize in Physics

*The Nobel Prize in Physics 2025 has been awarded to three researchers: the British-born physicist John Clarke, the American experimenter John Martinis, and the French theoretician Michel Devoret "for the discovery of macroscopic quantum mechanical tunnelling and energy quantisation in an electric circuit." As stated in the press release of the Nobel Committee, "this year's Nobel Prize laureates conducted experiments with an electrical circuit in which they demonstrated both quantum mechanical tunnelling and quantised energy levels in a system big enough to be held in the hand." Their achievements "open up possibilities for developing the next generation of quantum technologies, including quantum cryptography, quantum computers, and quantum sensors." This article places these discoveries in a historical context and highlights the role of earlier studies by other scientists — including researchers of the B.I. Verkin Institute for Low Temperature Physics and Engineering of the National Academy of Sciences of Ukraine, who obtained pioneering results in this field.*

*When big guys act like the little ones —
if you don't mess with them*

As always, each autumn, following the announcement of the Nobel Prize in Physics, the entire physical community actively discusses the nature of the research that led the laureates to this prestigious award. Professional scientists and popularizers of science strive to convey to the general public the significance of the results

PAPERS & REVIEWS

obtained and explain how they may influence the further development of science. In 2025, the Royal Swedish Academy of Sciences awarded the prize to two experimental physicists: the British-born John Clarke, who spent most of his career in the United States, and the American John Martinis, as well as to the French theoretical physicist Michel Devoret "for the discovery of macroscopic quantum tunneling and energy quantization in an electrical circuit."[1] Their key work was carried out in the mid-1980s at the University of California, Berkeley, which has produced 75 Nobel laureates. At that time, all three researchers worked together in John Clarke's group, in which John Martinis was his graduate student, and Michel Devoret joined as a postdoctoral researcher after defending his dissertation at the French Nuclear Research Center in Saclay.

The research of this year's Nobel laureates builds on earlier theoretical work by Anthony Leggett, who received the Nobel Prize in Physics in 2003 for his "pioneering contributions to the theory of superconductors and superfluids," but most of his subsequent work has focused specifically on quantum physics of macroscopic systems and condensed matter. The theoretical prediction of the possibility of macroscopic quantum tunneling is associated with Leggett's name [1], although Ukrainian theoretical physicists Yulii Ivanchenko and Lev Zil'berman from the Donetsk Institute for Physics and Engineering had already calculated the probability of such a process in a current-driven Josephson junction ten years earlier [2]. Their work was known to the aforementioned Nobel laureates and was even cited in the official scientific justification of the Nobel Committee[2].

Fascinating details of the Nobel laureates' scientific biographies, accounts of the sometimes-dramatic debates, and stories of numerous related scientific events involving both widely known and lesser-known researchers can be found in the memoirs and interviews of the laureates[3], published in various scientific journals and popular periodicals.

In the context of this series of works, it is appropriate to mention the chronology of discoveries in superconductivity, which is closely related to quantum mechanics, as well as to highlight the notable role played by the B.I. Verkin Institute for Low Temperature Physics and Engineering of the National Academy of Sciences of Ukraine (ILTPE NASU). Although for many physicists revisiting well-known facts about the development of superconductivity may seem redundant, this article is aimed at a broader scientific audience. Therefore, to provide a complete picture, it is worth first making a brief excursion into the prehistory of the works of this year's Nobel laureates, and only then examining their content and significance.

All the experiments discussed here were carried out on superconducting devices. As is well known, superconductivity is a purely quantum and collective phenomenon that cannot be understood without quantum mechanics. The main idea that enabled a microscopic description of superconductivity is the formation of paired electron states due to a weak effective attraction between them in a superconductor. The attraction is caused by phonon exchange, as theoretically shown by Herbert Fröhlich [3] and Leon N. Cooper [4] in the early 1950s. This two-electron formation is a boson with an integer spin, in contrast to the fermions (electrons) that constitute it. Such bosons tend to combine into a single whole — a superconducting Bose condensate, in which all particles occupy the same energy level and are described by a single wave function. Based on this concept, in 1957 John Bardeen, Leon N. Cooper, and John Robert Schrieffer formulated the so-called BCS theory [5], which considers many-particle interactions. For this work, they were awarded the Nobel Prize in Physics in 1972.

The next important milestone on this path was the theoretical prediction in 1962 by the young British researcher and Cambridge graduate Brian

---

[1] The Nobel Prize in Physics 2025. Press release. https://www.nobelprize.org/prizes/physics/2025/press-release/

[2] Scientific background to the Nobel Prize in Physics 2025. https://surl.lu/onoqnc

[3] For example, see First reactions. Telephone interview with John Clarke, October 2025. https://surl.lt/hmxzzf

Josephson of his two famous effects [6] (Nobel Prize in Physics, 1973).

The stationary Josephson effect, which involves the flow of a *superconducting* current through a thin dielectric tunnel barrier, unexpectedly revealed a large supercurrent density, comparable to the single-particle current density for the same barrier. This indicated the correlated tunneling of Cooper pairs as a single entity. In theoretical physics terms, this means that the tunneling probability of a correlated electron pair is proportional to the matrix element of the tunneling Hamiltonian, rather than to its square, as would be the case for uncorrelated electrons. Experimentally, Philip W. Anderson and John M. Rowell confirmed this effect the next year on a tin-lead tunnel junction [7], although other researchers, including the Ivar Giaever (whose Norwegian name is often mispronounced) and Hans Meissner, had previously observed the phenomenon in superconductor–oxide–superconductor and superconductor–normal metal–superconductor contacts, respectively.

Due to this effect, the phase of the superconducting condensate's wave function transformed from a theoretical concept into a physically measurable quantity (albeit still classical), since, according to Josephson's calculation, the superconducting current $I_s$ through a tunnel junction is proportional to the sine of the phase difference $\varphi$ between its two massive superconducting "banks": $I_s = I_c \sin\varphi$. In the context of the further discussion of macroscopic quantum tunneling, it should be noted that it is the tunneling probability of *individual* Cooper pairs, which are still microscopic objects, that is theoretically considered.

The second, non-stationary Josephson effect is less directly related to the main subject of our discussion, although it is arguably more important for practical applications. Moreover, this effect was the starting point of the long-term involvement of the young ILTPE in major international research of the superconductivity.

The non-stationary effect involves the emergence of an alternating current through a tunnel junction between two superconductors when a constant voltage is applied (this may sound straightforward to theorists, but experimentalists frown, since a constant voltage cannot be directly applied across a superconducting junction — it develops as the direct current through it exceeds the critical current $I_c$). The oscillation frequency $f$ of the current is strictly proportional to the applied constant voltage $V$, with the proportionality coefficient equal to the ratio of fundamental constants — twice the electron charge 2$e$ to Planck's constant $h$, which is approximately 483.6 MHz/μV:

$$f = \frac{2e}{h}V.$$

It is evident that an alternating current in a dipole, such as a Josephson junction, will lead to electromagnetic radiation from this "antenna". However, the power of such microwave (MW) emission into free space, even with impedance matching between the junction and the transmission line, is extremely low, measured in pico- or even femtowatts.

Indirect experimental confirmation of the second Josephson effect soon appeared in the form of so-called Shapiro steps on the current-voltage characteristics of the junction, which arise when it is irradiated by an external microwave field due to mixing with intrinsic Josephson oscillations [8]. Nevertheless, the primary interest lay in the direct detection of such radiation. Considering its extremely low power (~$10^{-12}$ W), the experiment to "capture" Josephson radiation was challenging. In this race for priority, a team of scientists from ILTPE — Igor K. Yanson, Igor M. Dmytrenko, and Vladimir M. Svistunov — took the lead. In 1965, they published their work [9] literally two months before a similar article appeared from an American group [10]. Remarkably, the Kharkiv team succeeded even though their competitors had far superior and more sensitive equipment. Subsequently, I.K. Yanson headed the tunnel microscopy department and, in collaboration with theorists, developed the method of point-contact spectroscopy [11]. For this reason, the department was later renamed by the method, and its staff further continued to work in this area.

It should be noted that the founder and first director of ILTPE, Boris I. Verkin, was not only a good and experienced scientist, but also an excellent and determined organizer. Due to his efforts many young and talented scientists joined the Institute's staff after 1960 when the institute was established, to make later significant contributions to global science. In this brief review of the history of so-called weak superconductivity (the future superconducting electronics), we will only briefly mention some of the achievements of ILTPE researchers, although the full list of accomplishments is, of course, much longer.

One example is the widely recognized theory of the Josephson effect in microbridges, which themselves are mesoscopic systems, i.e., intermediate between the micro- and macroscopic worlds. This theory, known as KO-1 and KO-2, was developed for two cases of electron motion — ballistic ("clean" bridges) and diffusive ("dirty" bridges) — by Igor O. Kulik and Alexander N. Omelyanchouk [12, 13]. The book by I.O. Kulik and I.K. Yanson "Josephson Effect in Superconducting Tunnel Structures" [14] has also achieved worldwide recognition.

The next significant step in the development of weak superconductivity, which is related to the 2025 Nobel Prize, involved two inventions. In 1964, a group of researchers at Ford Research Labs developed a superconducting direct-current quantum interferometer with two Josephson junctions (dc SQUID) [15], and then, in 1967, a single-junction radio frequency interferometer (rf SQUID) [16]. These devices became the "gold standard" in superconductivity physics and technology for many years, enabling the most sensitive measurements of weak magnetic fields in laboratory experiments and practical applications in medicine, geophysics, and other areas. Today, they form the basis for the creation of superconducting qubits, which are directly related to the topic under discussion. The first in the USSR dc SQUID was created at ILTPE already in 1967 (S. I. Bondarenko, I. M. Dmitrenko).

Regarding the rf SQUID, it principally is a small superconducting loop, only a few tens of microns in size, incorporating a Josephson junction. It directly links, through a proportionality, the phase difference $\varphi$ of the superconducting order parameter (the wave function of the superconducting condensate) at the ends of the junction to the magnetic flux $\Phi$ threading the loop: $\varphi = \Phi/\Phi_0$, where $\Phi_0$ is the so-called magnetic flux quantum. This flux (and therefore the phase) can be easily measured by classical methods.

As previously mentioned, A. Leggett once suggested the possibility of tunneling of quantum states as a single entity in macroscopic systems [1], considering the SQUID as such a system. From the point of view of radio physics, any electrically conductive object can be regarded as an antenna in the form of an electric or magnetic dipole, which can efficiently emit or absorb electromagnetic radiation if its size is comparable to half the wavelength. As the antenna size decreases, its efficiency rapidly drops, so microparticles hardly interact with photons.

In contrast, macroscopic systems fundamentally differ from microscopic objects in that they are strongly coupled to the electromagnetic noise environment, mostly of thermal origin, due to their relatively large size. Interaction with this noise reservoir leads to energy dissipation in the system, which destroys the coherence of possible macroscopic quantum states. This led to the introduction of new concepts into quantum mechanics, which originally did not account for dissipation and only considered reversible processes. Studying the tunneling of macroscopic states, Leggett laid the foundations for describing dissipative quantum processes in macroscopic systems [17, 18].

In this case, the tunneling probability depends on the energy dissipation, expressed as a certain generalized friction coefficient. Leggett and his collaborators paid special attention to the SQUID with magnetic flux trapped in its loop as a macroscopic quantum variable, seeing in it the most promising system for experimental verification of the theory. Damping in a macroscopic quantum system leads to the decay of coherent states, creating one of the most difficult challenges for experimentalists — the problem of isolating the quantum system from the electromagnetic environment. However, it is possible to measure the average decay rate as a function of temperature. This rate ceases to change as the temperature decreases, when thermally induced decays "freeze out," and quantum decays caused by macroscopic quantum tunneling (MQT) become observable.

Several research groups followed this approach in their experiments, attempting to observe the process of macroscopic quantum tunneling (MQT) in SQUIDs. Among them were ILTPE staff members Georgiy M. Tsoi and Vladimir I. Shnyrkov, who, under the supervision of Igor M. Dmitrenko, observed MQT in the superconducting loop of an rf SQUID in 1981 [19], grounding on the aforementioned ideas of A. Leggett [1, 17, 18] as well as the study by Yu.M. Ivanchenko and L.A. Zilberman [2]. Somewhat later, the talented physicist Viktor A. Khlus joined the team, providing theoretical support for these experiments. Their joint work [20], published in the ILTPE-issued journal Fizyka Nyzkykh Temperatur/*Low Temperature Physics*, is cited by this year's Nobel laureates in their "key" article [21]. We will briefly discuss [21] below and compare the different experimental approaches.

In [19, 20], it was shown that the frequency of spontaneous changes in the SQUID loop's magnetic state becomes temperature-independent at sufficiently low temperatures (0.5 K). This fact may indicate tunneling of magnetic flux (or the phase difference across the Josephson junction), reflecting the coherent behavior of a large number of correlated particles as a single entity with a macroscopically large total mass.

It should be noted that solution of the abovementioned problem of isolation from the external noisy environment that assumed meticulously designed filters, shields, transmission lines, and electronics, also required from researchers to make their experiments at night hours, when the TV center and public electric transport were inactive to eliminate extra interference. Only during this nighttime period could statistical measurement data be accumulated over several hours.

In fact, many researchers observed the MQT phenomenon (the most known works are [22—25]). They mainly used two types of superconducting macroscopic systems: an rf SQUID, in which magnetic flux piercing the loop tunneled between quantized current states, and a stand-alone Josephson junction with its tunneling superconducting condensate wave-function phase. Both approaches had their pros and cons. For example, a single-junction rf SQUID in the hysteretic regime has two or more local minima of potential energy (wells) depending on the internal magnetic flux. By applying an external flux, one can change the potential shape, thereby controlling the height of the potential barrier between neighboring wells. A barrier with a suitably small height (or more precisely, area) can result in a sufficiently high decay rate of the quantum states of the SQUID loop (i.e., change in the magnetic flux trapped in the loop) due to tunneling of the flux between wells, rather than caused by thermal fluctuations. A clear advantage of the single-junction rf SQUID over a stand-alone Josephson junction is the dissipation-free nature of these transitions: during phase tunneling, the Josephson junction in the loop does not enter a resistive state and therefore does not release heat, preserving the low temperature required for the experiment. This is exactly how the experiments were conducted at ILTPE.

This year's Nobel laureates were aware of these previous experiments but chose the approach involving a Josephson junction, which has the drawback mentioned above. However, as we will see, this was a deliberate choice, allowing them to definitively resolve the MQT problem. Recall that, according to Josephson's work, the potential of a tunnel junction periodically depends on the phase difference of the condensate wave function across it. When a current is applied, the potential tilts and takes the form of a "washboard" with local minima, the barrier height between which can be adjusted via the current. Each transition from one well to another is accompanied by a short voltage pulse, and therefore by energy dissipation and heating.

In their main experiment [21], John Clarke and John Martinis cooled a Josephson junction with a critical current of approximately 10 µA down to about 20 mK and applied short millisecond current pulses slightly below the critical current, measuring the average repetition rate of the voltage pulses that occurred during each current pulse. This rate corresponded to the inter-well transition rate and ranged from $10^{-2}$ to $10^{-6}$ s$^{-1}$. The amplitude of the current pulses had to be extremely stable, since the barrier height depends exponentially on the difference between the pulse amplitude and the critical current. At such a low temperature, the thermal time constant of the sample holder with the sample was so large (or "long," since it had dimension of time) that thermalization occurred slowly—even short current pulses had to be applied at a low repetition rate of 4—20 pulses per second. To obtain reliable statistics, $10^5$—$10^6$ voltage pulses, which emerged during each current pulse, had to be collected.

To create the required conditions, constrained by theoretical calculations, it was necessary to fabricate a tunnel junction with precisely specified parameters such as critical current, capacitance, and normal resistance. Next, the problem of isolation of the junction from external interferences must be solved. In addition to a double permalloy magnetic shield, the researchers installed a series of conventional RC-filter attenuators and developed a new type of coaxial microwave broadband powder filters to block the thermal irradiation from outside. Without exaggeration, the experimental setup became a true piece of art. It can be said that John Martinis contributed significantly to the preparation and successful execution of these experiments, thanks to his excellent skill as an experimental physicist.

It is important to explain why this work became a historic *experimentum crucis* in the study of MQT. In other experiments, the flattening of the interwell transition rate curve showing a plateau with decreasing temperature was considered evidence of reaching the quantum regime. However, temperature independence of the transition rate can also be caused by residual background noise that could not be completely eliminated. Therefore, despite qualitative agreement with theory, previous results were not definitive, although they could suggest MQT. Clarke and Martinis succeeded in independent measurements, made with sufficient precision using classical methods, the system

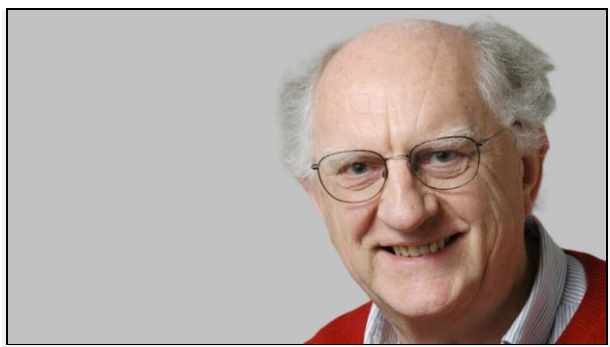

Photo: University of California, Berkeley

*John Clarke —*
*British experimental physicist, Professor Emeritus at the University of California, Berkeley (USA).*
*John Clarke was born on February 10, 1942, in Cambridge, United Kingdom. He studied at the Faculty of Natural Sciences at Christ's College, Cambridge, and earned a Bachelor's degree in Physics there in 1964. Then he worked at the Mond Laboratory of the Royal Society at the University of Cambridge. Clarke received his PhD in 1968 from Darwin College, where his doctoral advisor was the renowned physicist Brian Pippard. He repeatedly noted later that the greatest influence on his development as a scientist was exerted by another, elder student of B. Pippard, the 1973 Nobel laureate Brian Josephson, who predicted the Josephson effect in 1962. Subsequently, Clarke moved as a postdoc to the University of California, Berkeley, where he spent his entire academic career, holding positions as Assistant Professor (1969), Associate Professor (1971), and Professor of Physics (1973–2010).*
*Clarke's scientific interests focus on superconductivity and superconducting electronics, in particular the development of superconducting quantum interference devices (SQUIDs), which are ultra-sensitive detectors of magnetic flux. He worked on applications of SQUIDs configured as quantum-limited amplifiers for reading out superconducting qubits, new schemes for ultra-low-field NMR and magnetic resonance imaging, and the search for axions as a possible component of dark matter.*
*John Clarke is a member of the American Physical Society (1985), the Royal Society, London (1986), the U.S. National Academy of Sciences (2012), the American Academy of Arts and Sciences (2015), and the American Philosophical Society (2017). His awards include the Joseph F. Keithley Award for Advances in Measurement Science (1998), the Comstock Prize of the U.S. National Academy of Sciences (1999), the Hughes Medal (2004), the Olle W. Lounasmaa Memorial Prize (2004), and the Micius Quantum Prize (2021).*

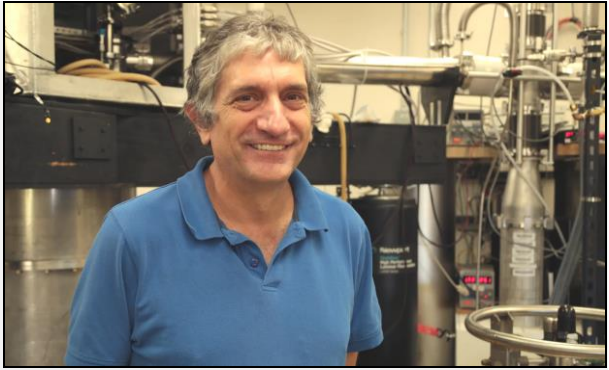

Photo: WIRED

*John Matthew Martinis —*
*American experimental physicist, Professor at the University of California, Santa Barbara (USA).*
*John Martinis was born in 1958 in San Pedro, California. His father, an ethnic Croat, immigrated to the United States to escape the communist regime in Yugoslavia.*
*Martinis graduated from the University of California, Berkeley, earning a Bachelor's degree in Physics in 1980 and a PhD in 1987 under the supervision of John Clarke. He then completed a postdoctoral fellowship in France at the Commissariat à l'Énergie Atomique in Saclay and later worked in the Electromagnetic Technologies Division at the U.S. National Institute of Standards and Technology, where he developed SQUIDs. From 2002, he began working on Josephson-junction qubits with the goal of creating the first quantum computer. His quantum device was recognized as the "Breakthrough of the Year" by Science magazine in 2010. In 2004, he joined the University of California, Santa Barbara, where he held the endowed Worster Chair in Experimental Physics until 2017.*
*In 2014, Google Quantum AI Lab signed a contract with Martinis' team to build a quantum computer using superconducting qubits. In October 2019, Nature published a paper in which Martinis' group reported achieving quantum supremacy for the first time using a 53-qubit quantum processor. In April 2020, Martinis left Google after being reassigned to a consulting position. He later moved to Australia to join the quantum computing startup Silicon Quantum Computing. In 2022, he founded Qolab, a company aimed at improving coherence in superconducting qubits to ensure more reliable and fault-tolerant quantum computing.*
*John Martinis is the recipient of the Fritz London Memorial Prize (2014) and the John Stewart Bell Prize for research on fundamental questions of quantum mechanics and their applications (2021).*

parameters required for theory: the Josephson junction's critical current, its capacitance, and, crucially, the losses, i.e., dissipation in the system. Choosing a single Josephson junction facilitated these direct measurements, providing an advantage over other approaches. After that, no fitting of parameters was needed, as comparison with theory was quantitative, and the measured average transition rates numerically matched the theoretical predictions.

Thus, the experiment demonstrated that a macroscopic degree of freedom (the phase difference across the Josephson junction, associated with the entire multi-particle superconducting condensate) obeys the laws of quantum mechanics. Further experiments involving microwave absorption (microwave spectroscopy) confirmed the existence of discrete quantized energy levels in this macroscopic system [26]. Overall, these results formed the basis for the Nobel Committee's decision to award the 2025 Physics Prize to John Clarke, John Martinis, and Michel Devoret for the discovery of macroscopic quantum tunneling and energy quantization in an electrical circuit.

The next stage in overcoming the challenges of practically harnessing the effects of quantum physics was associated with a more subtle but extremely important phenomenon — superposition of quantum states, which is even more remarkable in the case of macroscopic systems. It refers to the simultaneous existence of a system in two or more states and opens up the possibility of creating a fundamentally new type of information storage element — a qubit — as well as performing further operations with this quantum information. However, for these possibilities to be realized, it was necessary to reduce dissipation in a macroscopic quantum system by two to three orders of magnitude, which many considered impossible, unlike in the case of MQT. Researchers at ILTPE managed to observe this phenomenon earlier [27, 28] than others, but, as often happens in science, global priority was attributed to other researchers.

I would like to quote a passage from the memoirs of V.I. Shnyrkov (with his kind permission) about G.M. Tsoi, for a book written by his wife:

"In my opinion, the best results of Georgiy Mironovich Tsoi's research on SQUIDs were achieved in 1981—1983, when he studied macroscopic quantum tunneling (MQT), observed macroscopic resonant tunneling (MRT) and macroscopic quantum interference (MQI) (1983—1984). He also discovered and investigated (1985—1991) the phenomenon of coherent superposition of quantum states in macroscopic quantum oscillators (or the phenomenon of macroscopic quantum coherence, MQC). In the 2000s, such macroscopic quantum coherent states were called superconducting qubits (quantum bits). Unfortunately, for several years after the publication of our results on the discovery of MQC, some theorists from Moscow and Kharkiv, who at that time did not fully understand the complex properties of coherent quantum systems, criticized them. Thus, our results, indicating the discovery of qubits, were only published in conference proceedings… A fairly complete theory of the physical processes underlying flux qubits, experimentally observed by us in 1985—1991, was created only 15—20 years later."

Thus, the concept of superconducting qubits — phase, charge [29], and flux [30] ones — according to the official view, emerged in 1999. They laid the foundation for the subsequent rapid development of quantum engineering, with potential and already functioning applications. Those include quantum computers, ranging from the simplified quantum computer produced by D-Wave, which uses quantum annealing, to a whole series of quantum processors from Google, IBM, and other companies, as well as quantum-secured communication (successfully developed in many countries) and more speculative quantum radars. This progress enabled discussions of circuit quantum electrodynamics (cQED) and quantum computing algorithms.

In 2014, Google Quantum AI Lab invited John Martinis and his team, offering them a multimillion-dollar contract to develop a superconducting quantum computer. In 2019, they demonstrated the Sycamore processor with 53 qubits, achieving quantum supremacy over classical computers. However, in 2020, Martinis had to leave the company due to a conflict with management. In 2022,

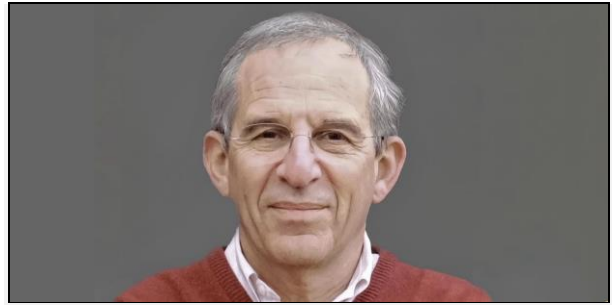

Photo: Yale University

***Michel Devoret —***
*French physicist, Professor at the University of California, Santa Barbara, Honorary Professor of Applied Physics at Yale University, and Chief Scientist at Google Quantum AI.*
*Michel Devoret was born in Paris, France, in 1953. He graduated from the École Nationale Supérieure des Télécommunications (ENST) in Paris with an engineer's degree in telecommunications. He earned a Master's degree in quantum optics at the University of Orsay, defended his dissertation in 1976 at the Laboratory of Molecular Photophysics at CNRS, and in 1982 obtained a PhD in condensed matter physics at the Atomic Energy Research Center in Saclay. From 1982 to 1984, Devoret worked as a postdoctoral researcher in John Clarke's group at the University of California, Berkeley. He then returned to France, where he founded the Quantronics group at the Orme des Merisiers laboratory of the Commissariat à l'Énergie Atomique in Saclay. There he studied tunneling times, invented the electronic pump, and developed a new type of qubit (the quantronium). In 2002, he became a professor at Yale University, where, together with his colleagues, he developed another type of superconducting charge qubit — the transmon. In 2009, he participated in the creation of a special type of flux qubit — the fluxonium. From 2007 to 2013, he worked at the Collège de France. In 2023, he was appointed Chief Scientist for Hardware at Google Quantum AI. In 2024, he joined the University of California, Santa Barbara, as a Professor of Physics.*
*Michel Devoret is a member of the American Academy of Arts and Sciences (2003), the French Academy of Sciences (2007), and the U.S. National Academy of Sciences (2023). He was knighted with the Legion of Honour in 2008. He has received numerous scientific awards, including the Ampère Prize of the French Academy of Sciences (1991), the Descartes-Huygens Prize of the Royal Netherlands Academy of Arts and Sciences (1995), the Europhysics-Agilent Prize of the European Physical Society (2004), the John Stewart Bell Prize (2013), the Fritz London Memorial Prize (2014), the Olle W. Lounasmaa Memorial Prize (2016), the Micius Quantum Prize (2021), and the Comstock Prize in Physics of the U.S. National Academy of Sciences (2024).*

he founded his private company, Qolab, which focused on superconducting quantum computing based on semiconductor chips technology.

It should be noted that even today, in these difficult times of war, ILTPE scientists remain engaged with major scientific trends, including quantum physics. For example, a group of young theorists led by Dr. S.M. Shevchenko actively develops the qubit topic. For two years, the Kharkiv Quantum Seminar, initiated jointly by ILTPE and the National Science Center "Kharkiv Institute of Physics and Technology," has been successfully operating, featuring lectures by well-known international scientists, including Nobel laureates. In recent years, experimental tasks in this field have become more challenging—what was a high achievement yesterday is nearly routine today. Conducting experiments now requires highly precise and expensive equipment like ultralow millikelvin temperature refrigerators, cryogenic amplifiers with quantum-limited sensitivity, specialized technological equipment, clean rooms, and specific materials. While awaiting improvements, particularly sufficient funding, our experimentalists are looking for opportunities to apply their knowledge and realize their ideas in international collaborative projects.

The 21st century is often called the quantum century, as the rapid development of technology, together with advances in theoretical and experimental physics in this field, has enabled the implementation of quantum ideas into real devices and new methods of information processing. Therefore, the Nobel Committee recognized not only outstanding achievements in fundamental physics but also the significant role they play in transforming everyday human life. It is a great pleasure to realize that the ILTPE scientists and Ukrainian researchers in general, were among the first to pave this path.

*This is an English translation of the original paper in Ukrainian made by the author (O. G. Turutanov). The page layout of the journal preprint retained, including headers, footers and page numbers, with minor changes. The author is grateful to the editors for providing a formatted preprint and biographies of the Nobel laureates.*

The original paper online: https://nasu-periodicals.org.ua/index.php/visnyk/article/view/25424/21677

This work and translation was supported by the project skQCI (101091548), funded by the European Union (DIGITAL) and the Recovery and Resilience Plan of the Slovak Republic.

**Cite the original article (in Ukrainian) in Visnyk of the National Academy of Sciences of Ukraine:**
Turutanov O.G. Nobel Prize and the contribution of Ukrainian scientists to the understanding of quantum phenomena, in particular the behavior of macroscopic quantum systems (Nobel Prize in Physics 2025). *Visn. Nac. Akad. Nauk Ukr.* 2025. (12): 20—30. https://doi.org/10.15407/visn2025.12.020